\documentclass[showpacs,floatfix,amsmath,twocolumn,amssymb,prb]{revtex4}

\usepackage{epsfig}

\begin{document}
 
\sloppy
 
\title{An experimental and computational investigation of structure and
magnetism in pyrite Co$_{1-x}$Fe$_x$S$_2$: Chemical bonding and 
half-metallicity}
 
\author{K. Ramesha} 
\email{ramesha@engineering.ucsb.edu}
\author{Ram Seshadri}
\email{seshadri@mrl.ucsb.edu}
\affiliation{Materials Department and Materials Research Laboratory\\
University of California, Santa Barbara, CA 93106 USA}
\author{Claude Ederer} 
\email{ederer@mrl.ucsb.edu}
\affiliation{Materials Research Laboratory\\
University of California, Santa Barbara, CA 93106 USA}
\author{Tao He} 
\email{tao.he@usa.dupont.com}
\author{M. A. Subramanian} 
\email{mas.subramanian@usa.dupont.com}
\affiliation{DuPont Central Research and Development\\
Experimental Station, E328/219 Wilmington, DE 19880-0328}

\begin{abstract}

Bulk samples of the pyrite chalcogenide solid solutions Co$_{1-x}$Fe$_x$S$_2$ 
$(0 \le x \le 0.5) $, have been prepared and their crystal structures and 
magnetic properties studied by X-ray diffraction and SQUID magnetization 
measurements. Across the solution series, the distance between sulfur atoms 
in the persulfide (S$_2^{2-}$) unit remains nearly constant. 
First principles electronic structure calculations using experimental crystal 
structures as inputs point to the importance of this constant S-S distance,
in helping antibonding S-S levels pin the Fermi energy. In contrast 
hypothetical rock-salt CoS is not a good half metal, despite being nearly 
isostructural and isoelectronic. We use our understanding of the 
Co$_{1-x}$Fe$_x$S$_2$ system to make some prescriptions for new ferromagnetic 
half-metals.

\end{abstract}


\maketitle

\section{Introduction}

The rapid development of spin valve-based magnetic read heads and the emergence
of spintronics\cite{spintronics} has thrown up a need for new magnetic 
half-metals for spin injection, as well 
as the need for a better understanding of the underlying materials issues in 
magnetic half-metals.\cite{coey_sanvito,pask} 
The recognition that pyrite CoS$_2$ 
is a ferromagnetic half-metal, \cite{jarrett,zhao} and that half-metallicity is
robust across the solid solution Co$_{1-x}$Fe$_x$S$_2$ \cite{mazin}
has led to considerable renewed efforts to understand this material.\cite{wang}
However, there is as yet no report on why the solid solution 
Co$_{1-x}$Fe$_x$S$_2$ is special: What are the unusual features in the crystal
and electronic structure of the pyrites that result in its properties ?

Benoit and N\'eel first showed that cobalt pyrite CoS$_2$ is a 
ferromagnet.\cite{neel} No other MX$_2$ compound (X = chalcogenide), or 
even MXY (Y = pnictide) is ferromagnetic.\cite{rao_pisharody,hulliger_mooser} 
Jarrett \textit{et al.}\cite{jarrett} made magnetic and transport measurements
on Co$_{1-x}$Fe$_x$S$_2$ which indicated itinerant electron ferromagnetism. 
FeS$_2$ ($x$ = 1) is a $d$ band semiconductor with filled octahedral 
$t_{2g}^6$ levels of Fe$^{2+}$ level separated from empty $e_g$ levels. 
As electrons are added ($0 \le x \le 1$) the compounds become conducting and 
ferromagnetic, even for $x$ values as large as 0.97 (or electron 
concentrations as small as 0.03 in the $e_g$ band). Over a wide range
of $x$, the magnetic moment (in Bohr magnetons) obtained from saturation 
magnetization is precisely equal to the number of $e_{g}$ electrons. 
DiTusa \textit{et al.}\cite{ditusa} have recently argued that the dilute ($x$ 
approaching 1) regions of the solid solution are worthy of closer examination 
and that near $x$ = 0.99, an insulator-metal transition is already observed. 
They report a quantum critical point in the ferromagnetic-paramagnetic 
transition between $x$ = 0.972 and $x$ = 0.964.

Spin-polarized electronic structure calculations by Zhao, Callaway and 
Hayashibara \cite{zhao} found that CoS$_2$ is ferromagnetic, and nearly a
half-metal, resembling the prototypic magnetic half-metal NiMnSb.\cite{deGroot}
Yamada \textit{et al.}\cite{yamada} have optimized the structure and Kwon
\textit{et al.}\cite{kwon} have performed LSDA+$U$ (LSDA = local spin density 
approximation) calculations on CoS$_2$. Shishidou 
\textit{et al.}\cite{shishidou} have performed first principles calculations 
on CoS$_2$ with gradient corrections (GGA = generalized gradient 
approximation). 

In a seminal paper, Mazin\cite{mazin} has shown from first-principles
calculations that ferromagnetic half-metallicity is ``robust'' in the system
Co$_{1-x}$Fe$_x$S$_2$, in the sense that in the region $0.85 \le x \le 0.25$ 
the compounds are perfect half metals, with moments precisely equal to the 
spin only values [$M(\mu_{\rm B})/$Co = 1] in agreement with the experiments 
of Jarrett \textit{et al.}\cite{jarrett} However, from point-contact Andreev 
reflection measurements, Cheng, Mazin and coworkers 
\textit{et al.}\cite{cheng} determine the maximum \textit{transport\/}
half-metallicity to not exceed 61\%. The maximum occurs near
$x$ = 0.5. The reduced half-metallicity is ascribed to sulfur deficiency 
in the samples, which interestingly, does not seem to affect magnetism.

In this contribution, we focus on the cobalt rich side of the pyrite
Co$_{1-x}$Fe$_x$S$_2$ phase diagram. We obtain a detailed structural 
description of the compounds $0 \le x \le 0.5$ from Rietveld\cite{rietveld} 
refinement of powder X-ray diffraction patterns. We also confirm from
magnetic measurements that the samples behave in the manner described by 
Jarrett \textit{et al.}\cite{jarrett} We use the crystal structures as inputs
for first principles electronic structure calculations based on the linear 
muffin-tin orbital method,\cite{andersen} both for pristine
CoS$_2$ as well as the supercells Co$_{0.75}$Fe$_{0.25}$S$_2$ and 
Co$_{0.5}$Fe$_{0.5}$S$_2$. We use the crystal orbital 
hamiltonian population (COHP)\cite{dronskowski} to examine details of 
spin-polarized chemical bonding across the solid solution series, and examine
the relation between chemical bonding and half-metallicity. A comparison with
rock-salt CoS (whose spin polarized crystal and electronic structure have been
calculated from first-principles) confirms the special features of the 
electronic structure of the pyrites.

%%%%%%%%%%%%%%%% Begin Figure %%%%%%%%%%%%%
\begin{figure}
\centering
\epsfig{file=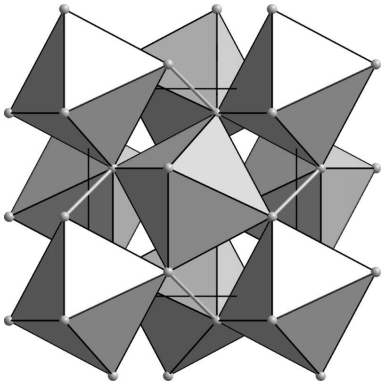, width=5cm}
\caption{MX$_2$ pyrite crystal structure showing MX$_6$ octahedra corner 
connected through X atoms which, in addition, are bonded (shown by 
sticks) to X atoms on neighboring octahedra. The coordination of X is 3(M) + 
1(X). X$_2$ sticks and M atoms (at the centers of the octahedra) form
two interpenetrating fcc lattices and a structure related to NaCl.}
\label{fig1}
\end{figure}
%%%%%%%%%%%%%%%% End Figure %%%%%%%%%%%%%%%

\section{Experimental}

Samples of Co$_{1-x}$Fe$_x$S$_2$ ($0 \le x \le 0.5$) were prepared starting 
from the elements taken according to stoichiometry, by heating well-ground 
powders in evacuated, sealed silica ampoules for 1 week at 673 K. The powders
were then reground, pelletized, resealed in evacuated silica ampoules, and 
heated for 873 K for 4 d. A final heating was performed at 973 K for one week,
of samples that had been ground up and pelletized again. Powder X-ray 
diffraction patterns were collected on powders using overnight runs on a 
Scintag X2 diffractometer operating in the Bragg-Brentano $\theta$-2$\theta$ 
geometry. Data were recorded using CuK$\alpha$ radiation and a step size of 
0.02$^{\circ}$ in 2$\theta$. The data were subject to Rietveld 
refinement\cite{rietveld} using the pyrite (space group $Pa\overline 3$, 
No. 205) structural model with the transition metal (Co or Fe) at $(0,0,0)$ and
S at $(x_{\rm S},x_{\rm S},x_{\rm S})$ with $x_{\rm S} \approx 0.39$. 
The \textsc{xnd}\cite{berar} Rietveld program was employed for the refinements.

Magnetic measurements were performed using a Quantum Design MPMS 5XL 
Magnetometer. Sample holders (gelatin capsules inserted in plastic 
drinking straws) held small solid pellets of the Co$_{1-x}$Fe$_x$S$_2$ phases.
We have not corrected the measured magnetizations for any core or 
sample-holder diamagnetism. Demagnetization corrections have not been 
performed.

\section{Computational methods}

Linear muffin-tin orbital (LMTO) calculations\cite{andersen} within the atomic 
sphere approximation (ASA) were performed using the 
\textsc{stuttgart tb-lmto-asa} program.\cite{stuttgart-LMTO}  Experimental
crystal structures used as inputs for the calculations were obtained from 
X-ray Rietveld refinements from this study, unless otherwise mentioned. 
Typically, more than 300 irreducible $k$ points within the primitive wedge 
of Brillouin zone were employed in the calculations. The generalized gradient 
approximation (GGA) for calculation of exchange correlation was employed 
following the Perdew-Wang prescription.\cite{perdew-wang} This results in 
slightly larger moments over the von Barth-Hedin\cite{vonBarth} LSDA, although 
not to the extent that CoS$_2$ is a perfect half metal as determined
by Shishidou \textit{et al.}\cite{shishidou} Calculations including the effect 
of the spin-orbit interaction were also performed using a modified version 
of the \textsc{lmto} code.\cite{Ederer_XMCD:2002} The implementation of the 
spin-orbit coupling into the otherwise scalar-relativistic LMTO formalism is 
analogous to the implementation described in \cite{MacDonald:1980} for the 
APW method. It was found that neither the states near the Fermi energy, not
the magnetic moment were in any way affected by the inclusion of spin-orbit 
coupling. For ferromagnetic, rock-salt CoS, the cell volume (which is the 
sole free structural parameter) was optimized using  full-potential 
linearized augmented plane wave (LAPW) calculations using the \textsc{wien2k}
code.\cite{wien2k} Exchange correlation was considered following the 
Perdew-Burke-Ernzerhof\cite{pbe} parametrization.

%%%%%%%%%%%%%%%% Begin Figure %%%%%%%%%%%%%
\begin{figure}
\medskip
\centering
\epsfig{file=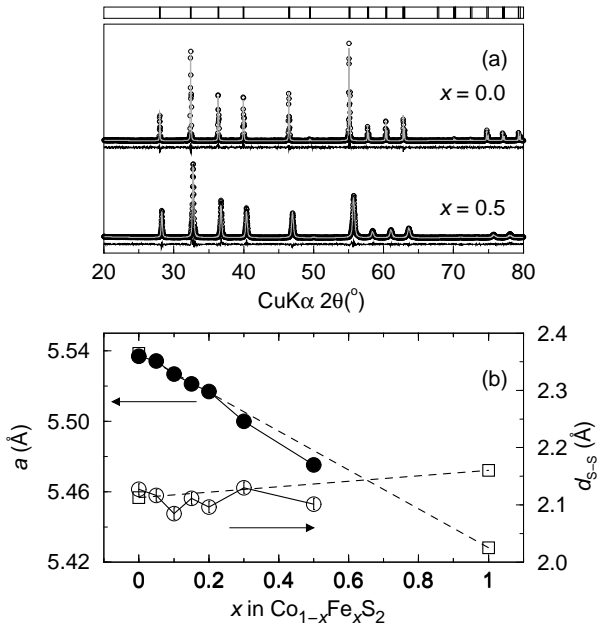, width=8cm}
\caption{(a) Powder X-ray Rietveld refinement of CoS$_2$ ($x$ = 0) and 
Co$_{0.5}$Fe$_{0.5}$S$_2$ ($x$ = 0.5). Data (circles), the Rietveld fit 
and the difference profiles are shown for each compound. Vertical 
lines at the top of the plot indicate expected peak positions. 
(b) Filled circles: Evolution of the $a$ cell parameter (in \AA\/) with $x$ 
of the solid solution Co$_{1-x}$Fe$_x$S$_2$. Error bars are smaller than 
the circles. The dashed lines connects published crystal 
structure\cite{nowack,finklea} data on the end members (squares). 
Open circles: S-S distances as function of $x$. The dashed line connects 
published\cite{nowack} data (squares).}
\label{fig2}
\end{figure}
%%%%%%%%%%%%%%%% End Figure %%%%%%%%%%%%%%%

\section{Results}

\subsection{Crystal structure}

Powder X-ray diffraction revealed all compounds in the series to be single 
phase, and well-fitted by Rietveld profile refinement to the pyrite crystal
structure described in FIG.~1. Results of the X-ray refinement are summarized
in FIG.~2(a), which shows data for the two extreme compositions [$x$ = 0.0 
and $x$ = 0.5] in the series studied here. The cubic $a$ cell parameter  
varies linearly with $x$, as shown in FIG.~2(b) indicating the formation of a 
homogeneous solid-solution. Careful analysis does however suggest a broadening 
in peak profiles as $x$ increases in Co$_{1-x}$Fe$_x$S$_2$. The decrease in the
$a$ lattice parameter as a function of increasing $x$ (substitution of Co by
Fe) arises from the different sizes of these ions; six-coordinate, low spin
Co$^{2+}$ has an ionic radius of 0.65~\AA\/ whereas the corresponding radius 
for Fe$^{2+}$ is 0.61~\AA.\cite{shannon} 
The single internal parameter in the pyrite crystal structure is the position 
$(x_{\rm S},x_{\rm S},x_{\rm S})$ of S. We have used refined values of 
$x_{\rm S}$ and $a$ to calculate S-S distances across the solid solution 
series. Within experimental error, we find nearly no change in the S-S 
distance as a function of $x$ as seen in FIG.~2(b). This is an important 
experimental observation, which we discuss at length at a later stage. 
In FIG.~2(b) we also show for comparison, structural data for the end-members 
CoS$_2$\cite{nowack} and FeS$_2$.\cite{finklea}

\subsection{Magnetism}

%%%%%%%%%%%%%%%% Begin Figure %%%%%%%%%%%%%
\begin{figure}
\medskip
\centering
\epsfig{file=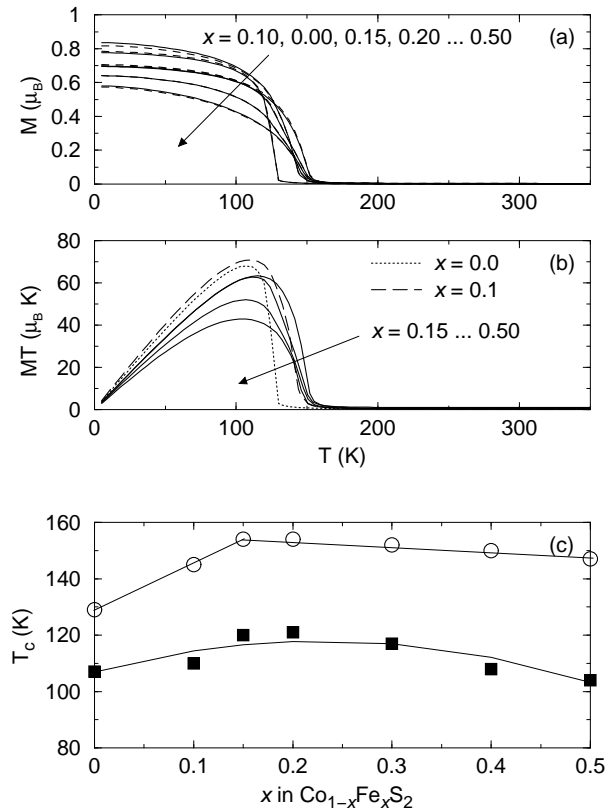, width=8cm}
\caption{(a) Zero-field cooled (dashed lines) and field-cooled magnetization
as a function of temperature of the Co$_{1-x}$Fe$_x$S$_2$ samples. 
(b) Field-cooled $MT$ \textit{vs.\/} $T$. (c) $T_c$ onset (circles) and 
midpoint, corresponding to the maximum value (squares) obtained from the 
$MT$ \textit{vs.\/} $T$ plot, as a function of $x$. The lines are guides 
to the eye.}
\label{fig3}
\end{figure}
%%%%%%%%%%%%%%%% End Figure %%%%%%%%%%%%%%%

Zero field cooled (ZFC) and field-cooled (FC) magnetization $M$ as a function 
of temperature recorded on Co$_{1-x}$Fe$_x$S$_2$ are indicated in FIG.~3(a). 
ZFC data were recorded in a field of 1000 Oe upon warming from 5 K after 
cooling from room temperature under zero field. FC data were also collected 
upon warming from 5 K, after the samples were cooled under a 1000 Oe field. 
All samples show evidence for ferromagnetism, with $T_c$s below 155 K. There 
is almost no ZFC/FC separation in any of the samples, suggesting the samples
are homogeneous, and also that they combine high permeability with low
saturation fields. Clear 
ferromagnetic $T_c$ onsets as well as widths of the transition are best seen 
from plots of $MT$ \textit{vs.\/} $T$ displayed in FIG.~3(b). The $T_c$ onset
does not seem to depend very much on $x$, and after an initially increasing
with $x$, almost remains constant as seen in FIG.~3(c). Data were acquired 
under relatively high field (1000 Oe) so even small clusters of spins are
sufficient for the magnetization to rise. The midpoints of the $MT$ 
\textit{vs.\/} $T$ traces are therefore better indication of ferromagnetic
$T_c$. These are also shown in FIG.~3(c), and are seen to initially increase
with $x$ and then decrease. The constant width of the transition [difference
between $T_c$ (onset) and $T_c$ (midpoint)] for the different values of $x$
reflects that all the samples are homogeneous, and the transition is not due
to small ferromagnetic clusters.

Magnetization at 5 K is displayed in FIG.~4(a). None of the samples showed any 
significant hysteresis implying Co$_{1-x}$Fe$_x$S$_2$ is a soft ferromagnet. 
Therefore only the positive $M$ \textit{vs.\/} $H$ quadrant is displayed.
All the samples display saturation at fields well below 1 T. The saturation
magnetization in Bohr magnetons ($\mu_{\rm B}$) is plotted as a function of
$x$ in FIG.~4(b). The dashed line is the expected spin-only value assuming 
each $e_g$ electron contributes 1 $\mu_{\rm B}$ per formula unit to the 
magnetization.  Only the parent CoS$_2$ phase is seen to have a saturation
magnetization less than the spin-only value. Starting from $x$ = 0.1 through 
$x$ = 0.5, all samples display spin-only behavior. This is an indication that 
all the samples except $x$ = 0.0 are within experimental error, perfect 
half-metals in terms of their being no ``leak'' in the magnetization from
majority to minority spin states. Such leaking is prevented by the complete 
absence of there are no minority spin states at the Fermi energy. Our results, 
for both ferromagnetic $T_c$ (midpoint) as well as saturation magnetization 
are nearly identical with those obtained by Jarrett 
\textit{et al.}\cite{jarrett}

%%%%%%%%%%%%%%%% Begin Figure %%%%%%%%%%%%%
\begin{figure}
\medskip
\centering
\epsfig{file=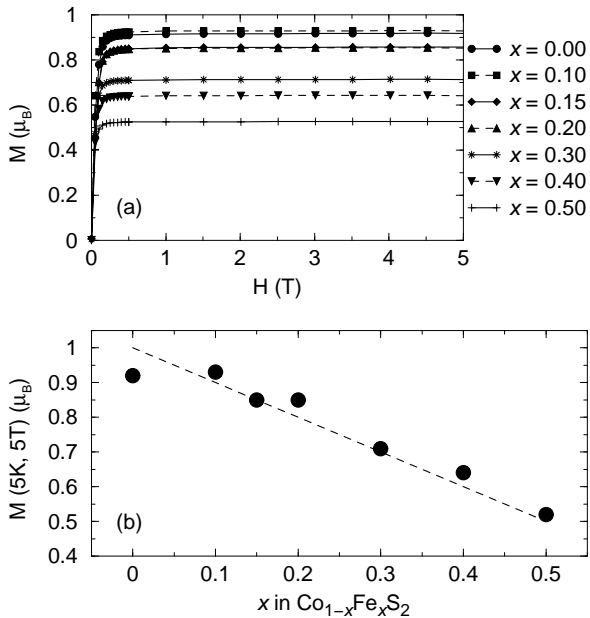, width=8cm}
\caption{(a) Magnetization at 5 K as a function of field. Since none of the
samples show appreciable hysteresis, only the positive quadrant is displayed.
Data were acquired from 5 T through 0 T. The dashed line is the expected
spin only value assuming every $e_g$ electron contributes 1 $\mu_{\rm B}$.
(b) Saturation magnetization (5 K, 5 T) as a function of $x$.}
\label{fig4}
\end{figure}
%%%%%%%%%%%%%%%% End Figure %%%%%%%%%%%%%%%

\subsection{Electronic structure}

%%%%%%%%%%%%%%%% Begin Figure %%%%%%%%%%%%%
\begin{figure}
\medskip
\centering
\epsfig{file=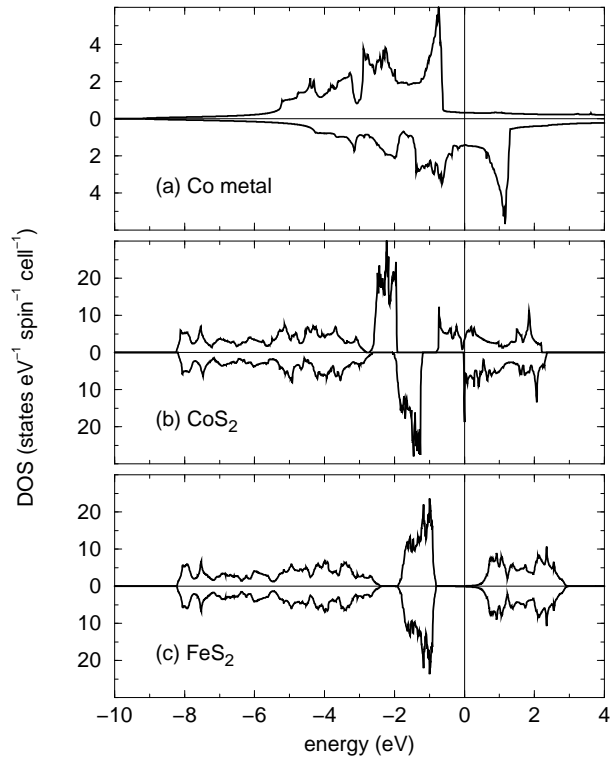, width=8cm}
\caption{(a) LMTO densities of state of \textit{hcp\/} Co metal. (b) Densities 
of state of CoS$_2$. The origin on the energy axis in (a) and (b), indicated 
by a vertical line, are the respective Fermi energies. (c) Densities of state 
of non-magnetic FeS$_2$ split into two spin directions. The energy axis has 
been shifted in (c) as described in the text. The upper and lower parts of 
each panel indicate respectively, majority and minority spin states.}
\label{fig5}
\end{figure}
%%%%%%%%%%%%%%%% End Figure %%%%%%%%%%%%%%%

An number of authors have provided detailed electronic structure descriptions
of CoS$_2$.\cite{zhao,yamada,kwon,shishidou} Mazin\cite{mazin} has examined 
magnetism across the series Co$_{1-x}$Fe$_x$S$_2$. The purpose of this section 
is to use structure refinements as inputs to obtain first principles 
electronic structures, and in particular, to calculate COHPs so that trends 
in spin-polarized bonding across the solid solution series can be obtained. 

Figure~5 shows total LMTO densities of state (DOS) in the two spin states for 
(a) \textit{hcp\/} Co metal, for (b) CoS$_2$ (using the $Pa\overline 3$ 
crystal structure obtained from X-ray refinements performed here) and 
(c) FeS$_2$ using the crystal structure reported by Finklea 
\textit{et al.}\cite{finklea}; SG. $Pa\overline 3$, $a$ = 5.4281 \AA, 
$x_{\rm S}$ = 0.38504. Fermi energies are taken as 0 on the energy axis in 
panels (a) and (b).  On going from Co metal to CoS$_2$, we observe a narrowing 
of $d$ states as well as the effects of the octahedral crystal generated by 
the $S_2^{2-}$ moieties. In Co metal, the Fermi energy lies in the minority 
spin states, in a region where majority [$s(\uparrow)$] states are also found.
Removal of these $s$ states by ionization (forming Co$^{2+}$ from Co)
is an essential ingredient in rendering the system half metallic. 

Panel (c) of this figure is the total DOS of non-magnetic semiconducting 
FeS$_2$ distributed equally between the two spin states. We have coincided 
$s$ states of S (in the region -20 eV to -10 eV with respect to the Fermi 
energy, not shown) by shifting the origin on the energy axis for the DOS of 
FeS$_2$, in order that the Fermi energy is fixed to the Fermi energy of 
CoS$_2$. The assumption is that S $s$ is a core state which should not be 
affected by compound formation. FeS$_2$ is a semiconductor with a calculated 
band gap of about 0.8 eV.\cite{eyert} In both FeS$_2$ and CoS$_2$, $p$ states 
of S below $E_{\rm F}$ extend from about -8 eV to -2.5 eV. In CoS$_2$, Co $d$ 
states (the $t_{2g}$ manifold) start at -2.5 eV where S $p$ states terminate. 
In FeS$_2$, there is a gap between occupied S $p$ states and the metal
$t_{2g}$ manifold. Comparing the DOS of FeS$_2$ with CoS$_2$, we observe that 
the $d$ manifold in the former is shifted to higher energies.
This is indicative of the general trend amongst the first row transition 
metals that as one goes to the right (from Sc through Cu), metal $d$ levels 
are stabilized. To some extent, this trend is reflected in the Pauling 
electronegativities which are 1.83 for Fe and 1.88 for Co. It is the same
trend which shifts MX$_2$ crystal structures from being layered (with 
M$^{4+}$) to being three-dimensional (with M$^{2+}$) in a process referred
to as redox competition.\cite{redox-competition} In oxide materials, the 
descent of cation $d$ levels as one traverses first row transition metals 
results in the famous Zaanen-Sawatzky-Allen phase diagram.\cite{zsa} 
In making solid solutions of CoS$_2$ and FeS$_2$, we believe the distinctly
shifted $d$ levels of FeS$_2$ have a role to play. While substitution of Co by
Fe in the series Co$_{1-x}$Fe$_x$S$_2$ results in electrons being removed from
the $e_g$ manifold, the $d$ levels themselves are pushed to higher energies;
the species (Fe) which ``removes'' electrons actually creates donor states. 
This is one of the factors which affects the electronic structure across the 
solid solution. A more electronegative substituent might remove electrons 
from $p$ states of S, and this would be disastrous for the magnetism as
demonstrated presently.

%%%%%%%%%%%%%%%% Begin Figure %%%%%%%%%%%%%
\begin{figure}
\medskip
\centering
\epsfig{file=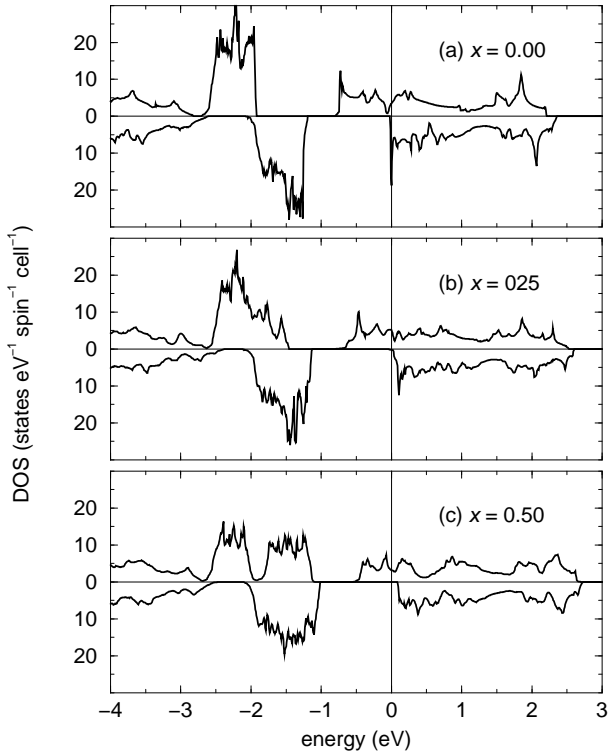, width=8cm}
\caption{Evolution of total LMTO densities of state as a function in $x$ 
in Co$_{1-x}$Fe$_x$S$_2$: (a) CoS$_2$ ($x$ = 0.00), 
(b) Co$_{0.75}$Fe$_{0.25}$S$_2$, ($x$ = 0.25), 
and (c) Co$_{0.50}$Fe$_{0.50}$S$_2$, ($x$ = 0.50).}
\label{fig6}
\end{figure}
%%%%%%%%%%%%%%%% End Figure %%%%%%%%%%%%%%%

We have performed LMTO calculations on ordered supercells of pyrite CoS$_2$ 
after systematically replacing some of the Co by Fe. Lattice $a$ and internal
structural parameters $x_{\rm S}$ for the calculations were taken from 
structure refinements of the nearest compositions as summarized in FIG.~2(b). 
In FIG.~6 shows densities of state for Co$_{1-x}$Fe$_x$S$_2$, for (a) 
$x$ = 0.00, (b) $x$ = 0.25, and (c) $x$ = 0.50.  In all three compounds, the 
shape of unfilled states just above $E_{\rm F}$ is ``box-like'' rising sharply 
with energy.  The evolution of $t_{2g}$ states with Fe substitution (in both 
spin directions) seems to result from a weighted superposition of the $t_{2g}$ 
states of spin-polarized CoS$_2$ and non-magnetic FeS$_2$ [shown in FIG.~5(b 
and c)]. Filled $t_{2g}$ levels below $E_{\rm F}$ seem pinned firmly in place. 
Partially filled $e_g$ levels are shifted up in energy, to near (the 
respective) Fermi energies.\cite{E_F_note} A feature of note
is that at $E_{\rm F}$, (the few) states in the minority states are 
progressively removed as $x$ increases in Co$_{1-x}$Fe$_x$S$_2$. This result,
as previously reported in the calculations of Mazin,\cite{mazin} explains
the less-than-perfect [ $(M/\mu_{\rm B})/\mbox{Co} < 1$ ] saturation 
magnetization of pure CoS$_2$ ($x$ = 0), and the increased 
[ $(M/\mu_{\rm B})/\mbox{Co} = 1$ ] magnetization as $x$ increases, seen in 
our magnetic measurements, and in the measurements of Jarrett 
\textit{et al.}\cite{jarrett} From a magnetism viewpoint, the extent of 
half-metallicity in this system can be obtained as the ratio of the saturation 
magnetic moment in Bohr magnetons to the number of unpaired $e_g$ electrons. 
Computationally, the magnetic moment, or more precisely, the polarization 
index $P$ given by:\cite{coey} 

\begin{displaymath}
P = \left| {\frac{N_\uparrow(E_{\rm F}) - N_\downarrow(E_{\rm F})}
            {N_\uparrow(E_{\rm F}) + N_\downarrow(E_{\rm F})}} \right|
\end{displaymath}

\noindent provides an indication of the half-metallicity. We calculate $P = 1$
for both the $x$ = 0.25 and the $x$ = 0.5 compounds. Correspondingly,
magnetic moments \textit{per\/} formula units were respectively obtained to be 
0.748 $\mu_{\rm B}$ and 0.500 $\mu_{\rm B}$; whereas for CoS$_2$ ($x$ = 0) 
it was 0.898 $\mu_{\rm B}$.

%%%%%%%%%%%%%%%% Begin Figure %%%%%%%%%%%%%
\begin{figure}
\medskip
\centering
\epsfig{file=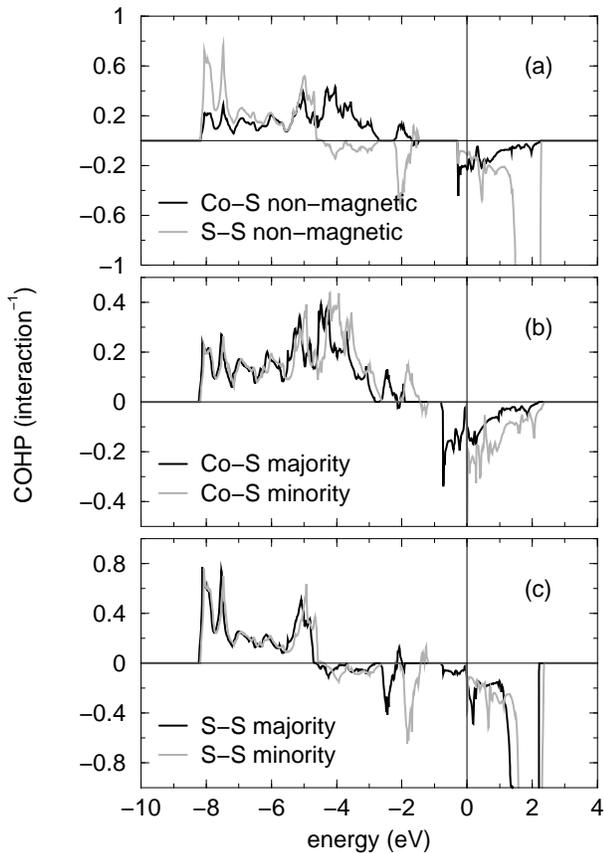, width=8cm}
\caption{(a) LMTO COHPs of non-magnetic CoS$_2$ showing Co-S and S-S 
interactions. The non-magnetic COHP has been scaled by a factor of 0.5.
(b) LMTO COHPs of the Co-S interactions in CoS$_2$ in the two spin directions.
(c) LMTO COHPs of the S-S interactions in CoS$_2$ in the two spin directions.
In the definition we employ here, positive COHPs correspond to bonding 
interactions and negative COHPs to antibonding interactions. This
is the opposite to the convention used in the original 
paper.\cite{dronskowski} 
        }
\label{fig7}
\end{figure}
%%%%%%%%%%%%%%%% End Figure %%%%%%%%%%%%%%%

The COHP\cite{dronskowski} is a very useful tool for mapping the energy 
dependence of pairwise bonding and antibonding interactions between atoms from
first-principles electronic structure calculations, including in systems
which are spin-polarized.\cite{landrum,felser} Figure~7(a) shows pairwise Co-S 
and S-S COHPs of parent \textit{non-magnetic\/} CoS$_2$, scaled by 0.5. 
We have verified that the spin-orbit coupling is negligible. Interactions are 
therefore confined to separate spin channels. Non-magnetic CoS$_2$ has
sharply antibonding states at the $E_{\rm F}$. Switching on spin-polarization
decreases these antibonding states, in keeping with the suggestion of Landrum 
and Dronskowski\cite{landrum} that sharply peaked antibonding COHPs in 
non-magnetic calculations can be an indicator (the equivalent of a Stoner 
criterion) of the electronic instability associated with spin polarization 
and ferromagnetism. 

From FIG.~7(b), we observe bonding Co-S COHPs in the region of $t_{2g}$ states 
and antibonding COHPs corresponding to the region of $e_g$ states. 
$E_{\rm F}$ in spin-polarized CoS$_2$ falls in a gap flanked by antibonding
Co-S($\uparrow$) and antibonding Co-S($\downarrow$), 
The S-S COHP in FIG.~7(c) shows the strongly bonding region where the $p$ 
states of S are found. The effect of spin-polarization on S-S COHPs is small
but important. Interestingly, the antibonding region of the S-S($\uparrow$) 
COHP just above $E_{\rm F}$ is slightly stabilized by spin-polarization, 
just as antibonding S-S($\downarrow$) is slightly destabilized. Antibonding 
S-S($\uparrow$) state are what pin the Fermi energy, and are perhaps the most 
significant states for discussing half-metallicity in these compounds.
S-S states are pseudo-molecular so they not disperse very greatly. They can be 
expected to remain in place because there is no great change in the charge 
state or in the degree of charge-transfer in the system as $x$ is increased,
as was observed from the constancy of the S-S distance. For antibonding 
S-S($\downarrow$) states to descend through the Fermi energy, the S-S bond 
would have to be elongated.

%%%%%%%%%%%%%%%% Begin Figure %%%%%%%%%%%%%
\begin{figure}
\medskip
\centering
\epsfig{file=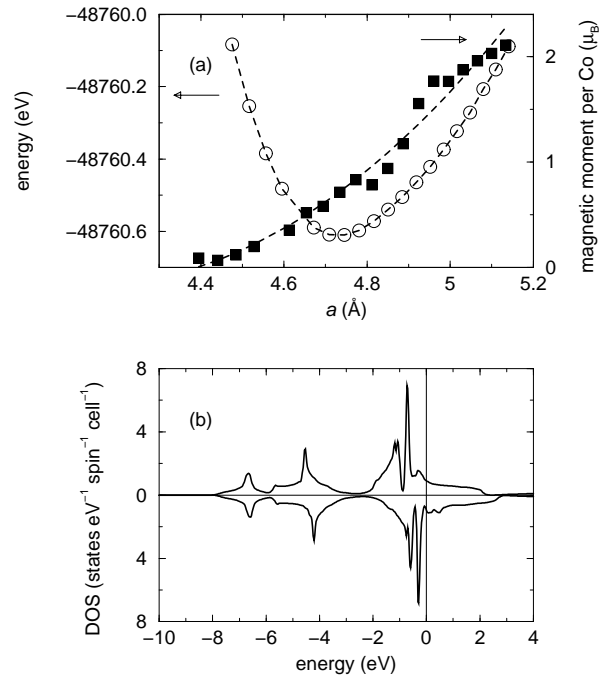, width=8cm}
\caption{(a) Total energy and magnetic moment \textit{per\/} Co atom in 
rock-salt CoS as a function of the cubic cell parameter, as obtained from 
spin-polarized LAPW calculations. (b) Total Densities of state in the two spin 
directions of ferromagnetic CoS, calculated for a rock salt ($Fm\overline 3m$)
structure with $a$ = 4.67 \AA.}
\label{fig8}
\end{figure}
%%%%%%%%%%%%%%%% End Figure %%%%%%%%%%%%%%%

%%%%%%%%%%%%%%%% Begin Figure %%%%%%%%%%%%%
\begin{figure}
\medskip
\centering
\epsfig{file=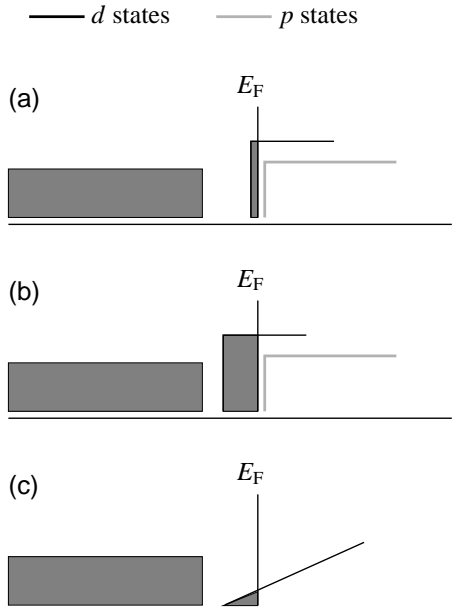, width=6cm}
\caption{(a) Summary of the non-magnetic electronic structure of 
Co$_{1-x}$Fe$_x$S$_2$ for large $x$ values (low $e_g$ filling).
As $x$ becomes slightly less than 1, the $e_g$ levels are filled and descend
below $E_{\rm F}$. The $p$ states remain above $E_{\rm F}$ however. The 
box-like shape of the unfilled $d$ states ensures the Stoner criterion is 
satisfied even for small filling. (b) Even at larger $e_g$ filling
(smaller $x$), only $d$ states descend below $E_{\rm F}$, and $E_{\rm F}$
is pinned to the bottom of the unfilled $p$ states. (c) Schematic non-magnetic
states in a more usual material such as CoS, where unfilled states grow 
gradually, and the Stoner criterion is satisfied only for large filling.
}
\label{fig9}
\end{figure}
%%%%%%%%%%%%%%%% End Figure %%%%%%%%%%%%%%%

In support of our argument that the S$_2^{2-}$ units play a crucial role in 
determining the electronic structure of CoS$_2$ and the series 
Co$_{1-x}$Fe$_x$S$_2$, we have performed first-principles calculations on
hypothetical rock-salt CoS, which has approximately the same atomic topology,
and the same formal Co valence as pyrite CoS$_2$. Figure~8(a) shows the results
of the structure optimization by plotting total energy as a function of the 
cubic cell parameter, as well as the corresponding magnetic moment of Co.
The GGA-optimized cell parameter was determined to be 4.67 \AA. The 
corresponding magnetic moment is about 0.5 $\mu_{\rm B}$ \textit{per\/} Co.
Figure.~8(b) shows the densities of state of ferromagnetic CoS in the 
two spin directions. It is seen that the crystal field in CoS is much 
smaller than in CoS$_2$. More importantly, CoS is not a magnetic half-metal,
despite minority spin states trying to nest in a pseudogap. The electronic
structure is characteristic of so-called ``intermediate spin'' systems such as
the finite-temperature electronic structure of the cobalt oxide perovskite 
LaCoO$_3$.\cite{imada} 

\section{Conclusions}

The low Curie temperatures of Co$_{1-x}$Fe$_x$S$_2$ make their use as spin 
injectors in spintronic circuitry unlikely.\cite{palmstrom} This system does 
however offer insights into the design of new half-metallic magnets. There
are two questions which our results help to address. The first is, what 
renders a compound half-metallic? From observing changes on going from 
ferromagnetic Co metal to CoS$_2$, we learn that $s$ and $p$ states at 
$E_{\rm F}$ (present in Co) are not good for half-metallic behavior since 
they are only poorly exchange split. Compound formation through removal of 
$s$ and $p$ electrons is therefore useful. This suggests that even in systems 
such as the Heusler compounds, X$_2$YZ, where X and Y are usually transition 
elements and Z is a main group element, it might help to have electronegative 
substituents at Z (for example, Si rather than Al).\cite{galanakis}

The second question is how the system retains ferromagnetism and 
half-metallicity across the substitution range. We summarize our findings on 
the unusual electronic structure of the pyrites solid solutions 
Co$_{1-x}$Fe$_x$S$_2$ in the scheme displayed in FIG.~9. For low filling
of $e_g$ states ($x$ approaching 1), the electronic structure is characterized 
by ``box-like'' states above $E_{\rm F}$, with a very sharp rise in the number 
of states with energy, as depicted in FIG.~9(a and b). The origins of this 
sharp rise, as we have demonstrated, are S-S antibonding states, which persist 
just above the $E_{\rm F}$ through the solid solution series. The states are 
sharp because they are pseudo-molecular. Even small filling of empty states 
results in the Stoner criterion being fulfilled\cite{mazin} and the rapid 
onset of ferromagnetism.\cite{jarrett,ditusa} The important role played by the 
shape of the DOS in fulfilling the Stoner criterion has been examined in 
detailed for transition metals by Andersen 
\textit{et al.}\cite{andersen_physica}

\acknowledgements{
We thank J. Gopalakrishnan for pointing us towards CoS$_2$ and Ryan Hummel
(IGERT undergraduate intern) for help with sample preparation. Discussions 
with, and help from Ole Andersen, Nicola Spaldin, Kiril Katsov, and 
Pio Baettig are gratefully acknowledged. This work was partially supported 
by the MRL program of the National Science Foundation under the Award 
No. DMR00-80034, including a seed grant. CE thanks Nicola Spaldin and
the Materials Research Laboratory for support.}

\clearpage


\begin{thebibliography}{99}
\bibitem{E_F_note} We find, using $s$ states of S as a reference, that the 
calculated position of the Fermi energy shifts only very little across the 
solid solution series.
\end{thebibliography}
\end{document}